\documentclass[a4paper,twoside,10pt]{letter}
\usepackage{graphicx,saj,multicol,subeqnarray}

\def\papertype{}
\def\dj{d\kern-0.4em\char"16\kern-0.1em}

\begin{document}
\baselineskip=3.1truemm
\columnsep=.5truecm
\newenvironment{lefteqnarray}{\arraycolsep=0pt\begin{eqnarray}}
{\end{eqnarray}\protect\aftergroup\ignorespaces}
\newenvironment{lefteqnarray*}{\arraycolsep=0pt\begin{eqnarray*}}
{\end{eqnarray*}\protect\aftergroup\ignorespaces}
\newenvironment{leftsubeqnarray}{\arraycolsep=0pt\begin{subeqnarray}}
{\end{subeqnarray}\protect\aftergroup\ignorespaces}
%


\markboth{\eightrm Newly Observed Small-Scale Structures in the Quiet Sun and Their Connection to Solar Oscillations}
{\eightrm A. Andic}

{\ }

\publ

\type

{\ }


\title{Newly Observed Small-Scale Structures in the Quiet Sun and Their Connection to Solar Oscillations}


\authors{A. Andic$^{1}$ }

\vskip3mm


\address{$^1$Big Bear Solar Observatory,NJIT, 40386 North Shore Lane, Big Bear City, CA 92314, USA}


\dates{November 23, 2009}{December 13, 2009}


\summary{Our observations of the quiet Sun with the NST have yielded unanticipated results on smal-scale solar dynamic. Althought small-scale solar dynamic have been well-studied, the NST is enabling us to probe finer scale dynamic exploiting higher spatial resolution.\\ We discuss NST resutls from data taken 29th July 2009 using an broadband filter centered on TiO 705.7nm spectral line. Data are from the center of the solar disk where we observed quiet Sun.\\ We registered bright-point like structures in most of the intergranular lanes. They vary in behaviour and evolution. Co-registered solar oscillations tend to congregate near certain types of bright-point like structures. The oscillations with maximum power tend to appear only above or near these structures.\\ Work is accepted and presented at AGU annual fall meeting 2009 in San Francisco, USA as poster. }



{


\section{1. Introduction}\label{intro}
 
The resolution of the 1.6 m NST at BBSO gave us the unique opportunity to observe previously unresolved structures. 
Previous research established that observed oscillatory behavior prefers inter-granular lanes in a seemingly stochastic manner. The resolution of  0.1 gave me opportunity to check this point.

\section{2. Data}

The data-set used for this work is part of observing sequence obtained on 29th July 2009. Nasymth optical setup  that  consists from TiO broadband filter and PCO camera is used at NST at BBSO. A filter has central wavelength at 705.68nm and bandpass 1nm. The quiet Sun at the centre of the disk is observed. The data sequence consist of 120 speckle reconstructed images with cadence of 15s. Speckle reconstruction was done with KISIP code [Woegler \it{et al.} 2008].

\section{3. Results}


\centerline{\includegraphics[height=6.5cm]{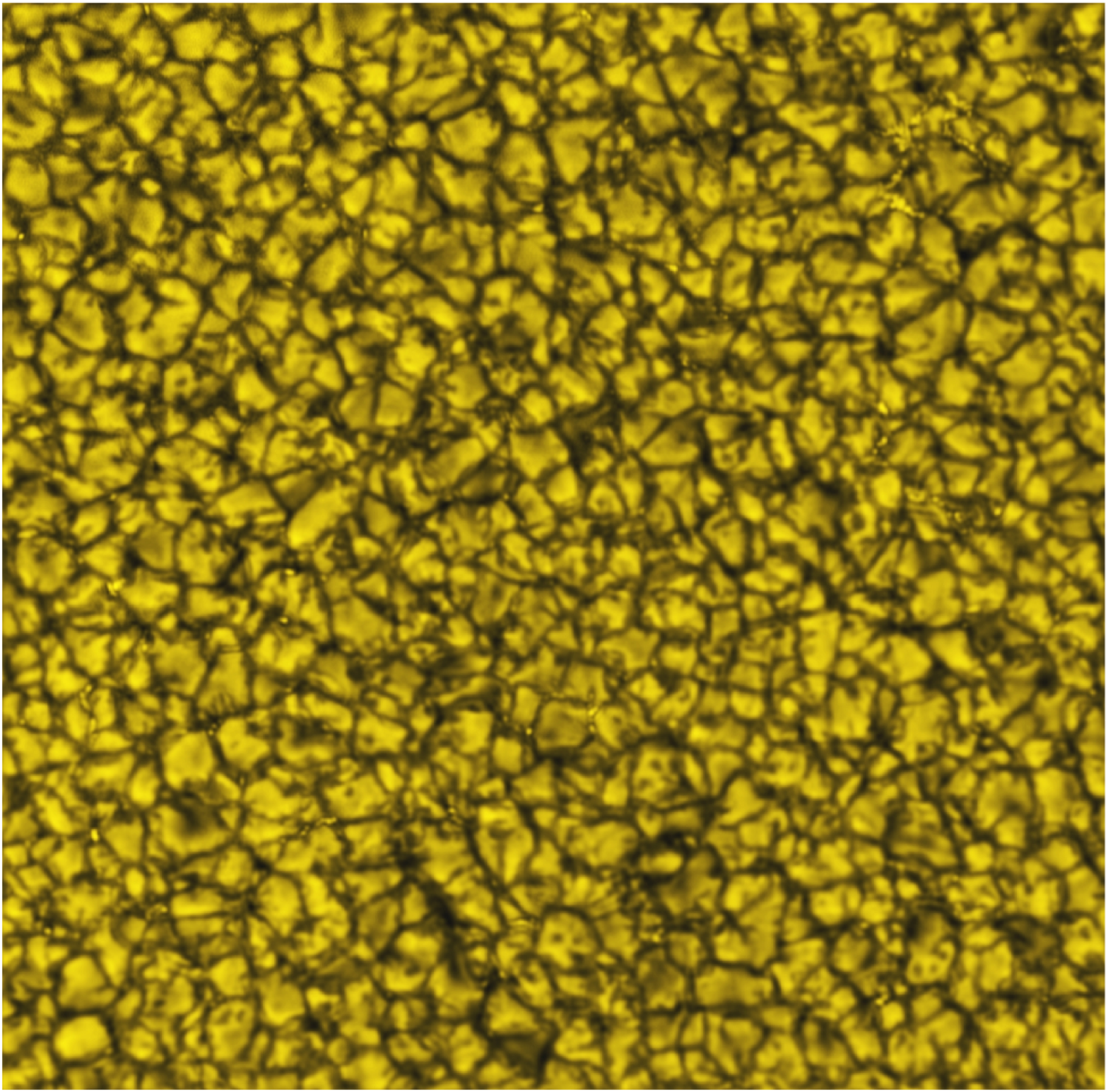}}


\figurecaption{1.}{A granulation image obtained with NST. Although the quiet Sun is observed the amount of the bright-point like structures in the inter-granular lanes is much larger than expected. Due to the molecular nature of the observed line we assume that it behaves in similar way as G-Band.}

\centerline{\includegraphics[height=6.5cm]{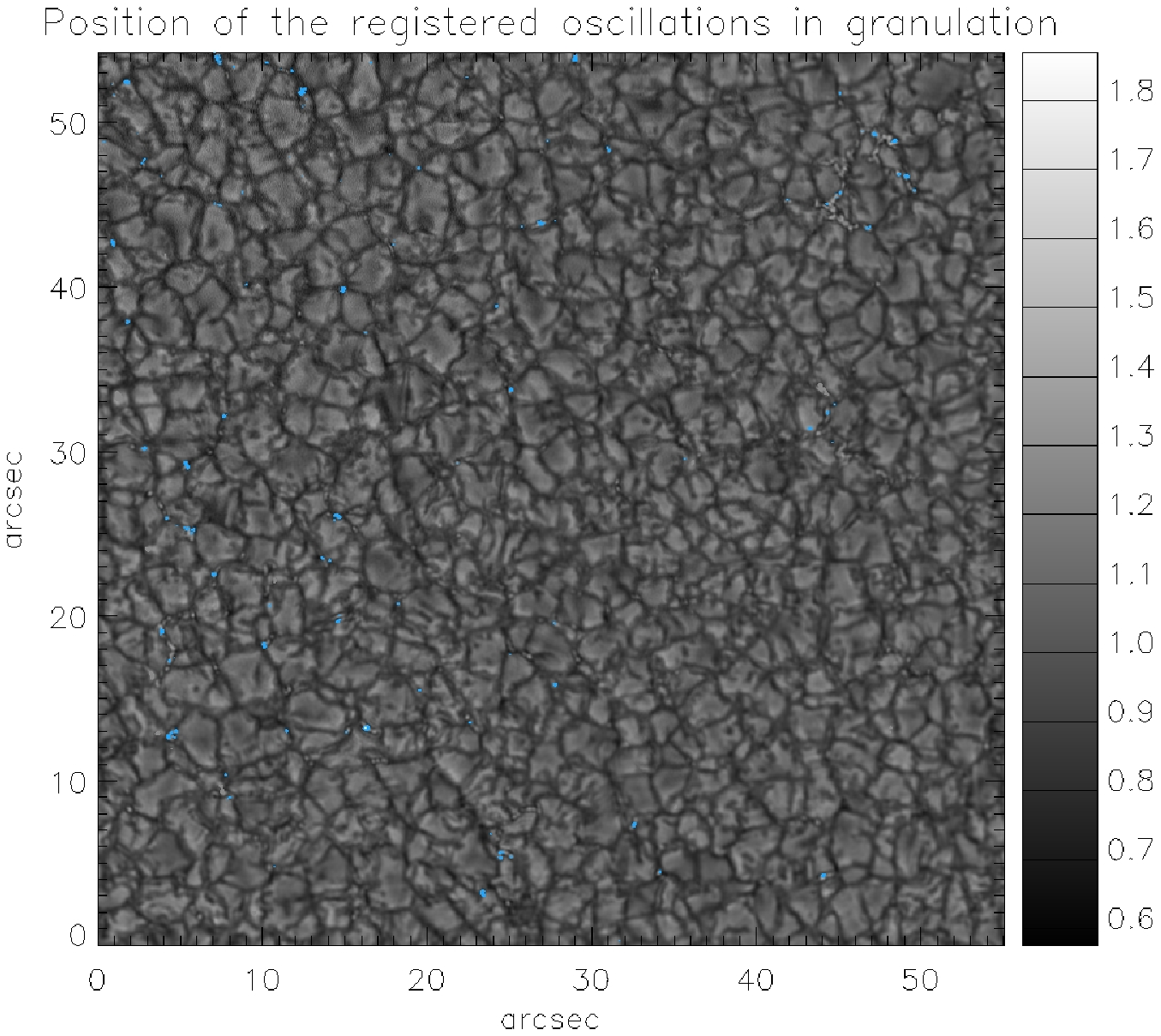}}


\figurecaption{2.}{The image shows the locations of the registered oscillatory power. The oscillatory power is represented with the blue contours. Shown is all of the registered oscillatory power above 80\% of the maximum registered power appears above the bright-point like structures or very close by. }

\centerline{\includegraphics[height=6.5cm]{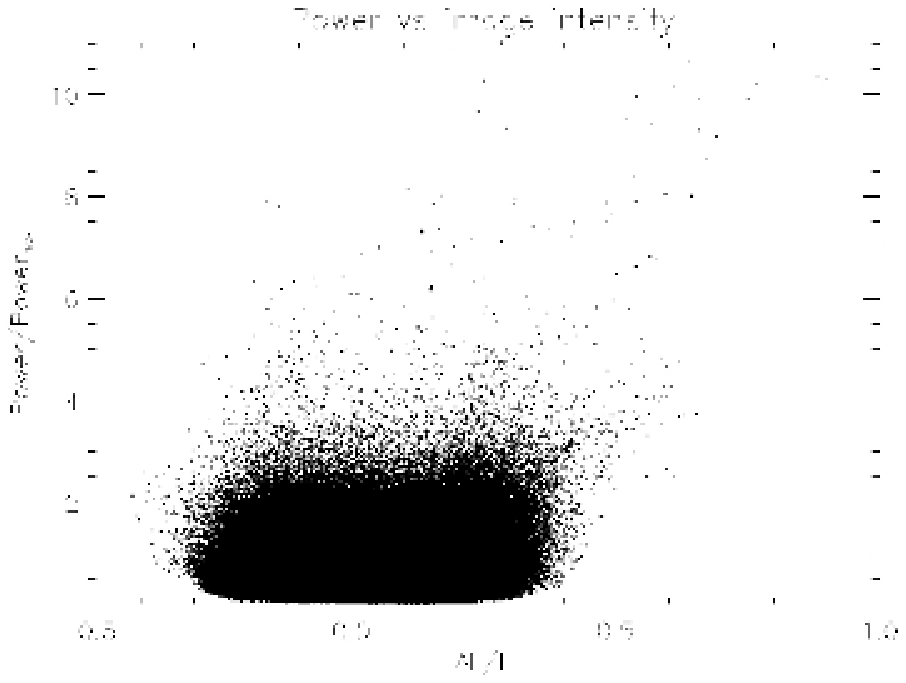}}


\figurecaption{3.}{The scatter plot presents the position of the all registered oscillatory power  vs image intensity. Opposite to findings of previous research, the majority of the observed power is located above the brighter structures, as demonstrated in image above, too.  }

\section{4. Analysis and  Conclusions}

The data-set was speckle reconstructed. Oscillations were registered using wavelet analysis with special care taken to pay attention only to the oscillations that were at least 95\% significant. The frequency range of analyzed oscillations was from 1.4mHz to 33mHz.\par

The position of the oscillations mostly above the bright-point like structures indicate the significant role of the magnetic field in either their creation or their propagation. 

\references

Woeger, F., von der Luehe, O., Reardon, K. 2008, \journal{A\&A},\vol{488},375W

\endreferences

}

\end{document}